\documentstyle[aaspp4,11pt]{article}

\def\gsim{\;\rlap{\lower 2.5pt
\hbox{$\sim$}}\raise 1.5pt\hbox{$>$}\;}
\def\lsim{\;\rlap{\lower 2.5pt
   \hbox{$\sim$}}\raise 1.5pt\hbox{$<$}\;}

\def\spose#1{\hbox to 0pt{#1\hss}}                
\def\lta{\mathrel{\spose{\lower 3pt\hbox{$\mathchar''218$}}
     \raise 2.0pt\hbox{$\mathchar''13C$}}}
\def\gta{\mathrel{\spose{\lower 3pt\hbox{$\mathchar''218$}}
     \raise 2.0pt\hbox{$\mathchar''13E$}}}
\newcommand{\beq}{\begin{equation}}
\newcommand{\eeq}{\end{equation}}



\lefthead{MENOU, PERNA \& HERNQUIST} 
\righthead{RADIO PULSAR SPINDOWN}

\begin{document}                                               
\title{Disk-Assisted Spindown of Young Radio Pulsars}

\author{Kristen Menou,\altaffilmark{1}}
 
\affil{Princeton University, Department of Astrophysical Sciences,
Princeton NJ 08544, USA, kristen@astro.princeton.edu}

\author{Rosalba Perna\altaffilmark{2} and Lars Hernquist}
 
\affil{Harvard-Smithsonian Center for Astrophysics, 60 Garden Street,
Cambridge MA 02138, USA, rperna@cfa.harvard.edu,
lars@cfa.harvard.edu}
 
\altaffiltext{1}{Chandra Fellow}
\altaffiltext{2}{Harvard Junior Fellow}

\begin{abstract}

We present a model for the spindown of young radio pulsars in which
the neutron star loses rotational energy not only by emitting magnetic
dipole radiation but also by torquing a surrounding disk
produced by supernova fallback.  The braking index predicted in our
model is in general less than $n=3$ (the value for pure dipole
magnetic radiation), in agreement with the reported values of $n<3$
for five young radio pulsars.  With an accuracy of $30\%$ or better,
our model reproduces the age, braking index and third frequency
derivative of the Crab pulsar for a disk mass inflow rate in the range
$3 \times 10^{16} - 10^{17}$~g~s$^{-1}$.

\end{abstract}

{\it subject headings}: accretion, accretion disks --
supernovae: general -- pulsars: general -- stars: neutron

\section{Introduction}

The magnetic dipole model for radio pulsars has proven extremely
successful in explaining the observational properties of these objects
(Pacini 1967; 1968; Gunn \& Ostriker 1969; Gold 1968; Goldreich \&
Julian 1969; see, e.g., Shapiro \& Teukolsky [1983] for an overview;
Manchester \& Taylor [1977] for a detailed review). A longstanding
problem of the theory, however, has been that the predicted braking
index (see \S2.1 for definition) is $n=3$, while measurements of braking
indices in five young radio pulsars are all less than 3 (see \S3 for
the most recent measured values).  Various effects have been proposed
to explain this discrepancy, such as magnetic axis wandering (Macy
1974), non-standard vacuum dipole theory (Melatos 1997), and
non-dipolar field structure or field strength secular evolution
(Blandford, Applegate \& Hernquist 1983; Blandford \& Romani 1988).

In this {\it Letter}, we explore the consequences of a recent proposal
by Chatterjee, Hernquist \& Narayan (2000) according to which young
radio pulsars could be surrounded by remnant disks produced by
supernova fallback (see also Michel \& Dessler 1981, 1983; Michel
1988; Yusifov et al. 1995; Alpar 1999, 2000).  The disk may indeed be
able to contribute to the neutron star spindown even in the radio
pulsar phase, provided that matter does not fall through the
magnetosphere inside the light cylinder radius.  We note that Marsden,
Lingenfelter \& Rothschild (2001a,b) have recently considered some of
the consequences of the presence of such a disk around a radio pulsar,
with a special emphasis on explaining discrepancies between a pulsar's
true age and its timing age (see also Gvaramadze [2001] for the case
of an interaction with dense circumstellar material).  The outline of
this {\it Letter} is as follows. In \S2, we derive a theory for pulsar
spindown in the presence of a fallback disk and the associated
propeller torques. We apply the theory to the Crab pulsar and other
young radio pulsars in \S3.

\section{Pulsar Spindown Model}

We certainly lack a proper understanding of the magnetospheric
interaction between a radio pulsar and a surrounding disk. However,
simple arguments can be used to derive a reasonable rate at which the
pulsar is spun down by this interaction. Implicit to our model is the
assumption that the disk can survive the high-energy environment
created by the intense relativistic pulsar wind. For the pulsar
mechanism to operate, we require the disk to be located beyond the
light cylinder radius, at $R_{lc}=c/\Omega$ (where $\Omega$ is the
neutron star angular speed and $c$ is the speed of light). On the
other hand, the magnetic field lines are open beyond $R_{lc}$ and the
field structure is no longer dipolar. This is the main motivation for
our assumption that the disk extends down to $R_{lc}$. At $R_{lc}$,
the inflowing material encounters a dipolar field structure rotating
at nearly the speed of light: an efficient propeller effect
(Illarionov \& Sunyaev 1975) may be expected to operate there,
spinning down the neutron star and preventing mass inside the pulsar
magnetosphere at the same time.\footnote{We note that, if the disk
were truncated beyond $R_{lc}$ (e.g. because of a large energy density
right outside the light cylinder), the torque on the neutron star
would likely be largely reduced or entirely suppressed.}

Following Menou et al. (1999; see also Daumerie 1996), 
we adopt an efficient propeller torque
\begin{equation}
\dot J \equiv I \dot \Omega = -2 \dot M R_m^2 \Omega,
\end{equation}
where $I$ is the moment of inertia of the neutron star, a dot denotes
a time derivative and $\dot M$ is the disk mass inflow rate at the
magnetospheric radius, $R_m$. This expression for the torque assumes
that the material flung away by the propeller effect has been
accelerated to an angular speed corresponding to that of the star. In
the pulsar case, we effectively take $R_m=R_{lc}$ and the material
flung away only approaches the speed of the light. Under these simple
assumptions, the torque for this mildly relativistic propeller is
given by
\begin{equation}
\dot J = -2 \dot M c^2 \Omega^{-1},
\end{equation}
which is essentially what one expects from dimensional arguments
assuming the torque acts at $R_{lc}$. A more efficient, relativistic
propeller is energetically possible; it would correspond to a reduced
value of the disk mass inflow rate for a given value of the torque.

The pulsar loses rotational energy at a rate:
\begin{equation}
\dot E \equiv I \Omega \dot \Omega = -\beta \Omega^4 - \gamma,
\label{eq:rotloss}
\end{equation}
where the first term on the RHS corresponds to magnetic dipole
radiation losses and the second term to propeller losses.  We have
\begin{equation}
\gamma = 2 \dot M c^2 \approx 2 \times 10^{37}~{\rm erg~s^{-1}} \left(
\frac{\dot M}{10^{16}~{\rm g~s^{-1}}}\right),
\end{equation}
and
\begin{equation}
\beta = \frac{B_p^2 \sin^2 \alpha R_{NS}^6}{6 c^3},
\end{equation}
where $R_{NS}$ is the neutron star radius and $B_p$ is the magnetic
field strength at the magnetic pole, inclined at an angle $\alpha$
from the rotation axis (see, e.g., Shapiro \& Teukolsky 1983).

In what follows, we assume that Eq.~(\ref{eq:rotloss}) is valid at all
times. We note that this assumption is not entirely consistent with
the evolutionary scenario of Chatterjee et al. (2000), in which a
neutron star decouples completely from its surrounding fallback disk
when it enters the radio pulsar phase. We expect this discrepancy to
be minor given that magnetic dipole radiation still dominates the
losses in the present model. More importantly, a young neutron star is
not initially in the radio pulsar phase in the Chatterjee et
al. model, but in a ``dim propeller'' phase (during which the
magnetospheric radius has not yet reached the light cylinder
radius). This phase can last up to several hundreds of years in the
Chatterjee et al. model and it is not captured by our
Eq.~(\ref{eq:rotloss}). This discrepancy suggests that the results
derived below for the integrated neutron star spin history should be
interpreted with caution (instantaneous quantities such as the braking
index are not directly affected, however).

\subsection{Steady mass inflow rate}

{}From Eq.~(\ref{eq:rotloss}), we find that the braking index is
\begin{equation}
n \equiv \frac{\ddot \Omega \Omega}{\dot \Omega ^2} = 3 -
\frac{1}{\frac{1}{4}+\frac{\beta}{4 \gamma} \Omega^4},
\label{eq:nsteady}
\end{equation}
if $\gamma$ (i.e. the disk mass inflow rate) is constant with time. The
breaking index is therefore less than $3$, the value
corresponding to pure magnetic dipole radiation losses (recovered in
the limit $\gamma \rightarrow 0$).

Following the usual procedure, Eq.~(\ref{eq:rotloss}) can be separated
and integrated to constrain the neutron star spin history. We rewrite
Eq.~(\ref{eq:rotloss}) as
\begin{equation}
\frac{d \Omega ^2}{1+\frac{\beta}{\gamma} \Omega^4}=-\frac{2 \gamma}{I}dt,
\end{equation}
which can be integrated (assuming $\gamma=$~constant) by noting that
the LHS is of the form $dy/(1+y^2)= d(\arctan(y))$.  This yields, at
any time $t$ after the system's birth:
\begin{equation}
t=\frac{I}{2(\gamma \beta)^{1/2}} \left[ \arctan\left[ \left(
\frac{\beta}{\gamma} \right)^{1/2} \Omega_i^2\right] - \arctan \left[
\left( \frac{\beta}{\gamma} \right)^{1/2} \Omega^2(t)\right]\right],
\label{eq:trueagesteady}
\end{equation}
where $\Omega(t)$ is the angular rotation speed at time $t$ and
$\Omega_i$ is the angular rotation speed at birth. In the limit of
very large $\Omega_i$, the left $\arctan$ is simply equal to $\pi/2$.
There is no simple relation between the pulsar timing age and
true age in this solution, contrary to the case of pure magnetic
dipole radiation losses (Shapiro \& Teukolsky 1983; Gunn \& Ostriker
1969).

\subsection{Time-dependent mass inflow rate}

We generalize here the above calculations to the case of a mass inflow rate
decreasing with time, as expected in the fallback disk scenario
(Chatterjee et al. 2000; Menou, Perna \& Hernquist 2001). Allowing the mass inflow rate
to vary with time, the braking index becomes:
\begin{equation}
n  = 3 - \frac{1 - \frac{1}{4} \frac{\dot \gamma}{\gamma}
\frac{\Omega}{\dot \Omega}}{\frac{1}{4}+\frac{\beta}{4 \gamma}
\Omega^4},
\label{eq:nunsteady}
\end{equation}
where both $\dot \gamma$ and $\dot \Omega$ are negative (assuming a
decreasing mass inflow rate).\footnote{We note that this expression for
the braking index can potentially yield values of $n >3$.} The
quantity $- \Omega / \dot \Omega$ is twice the usual pulsar timing age
($\equiv P/2 \dot P$), while the quantity $- \alpha \gamma / \dot
\gamma$ is actually the true age of the system if, as expected, the
mass inflow rate obeys a power law evolution in time of the form
$\gamma(t)\propto t^{-\alpha}$.

While the braking index and the second deceleration parameter (defined
in \S 3.1) are instantaneous quantities which can easily be computed
for any power-law solution (using, if known, the values of the true
age and the timing age as explained above), the spin history cannot be
computed analytically for an arbitrary power law index $\alpha$, but
only in some special cases.

For simplicity, here we restrict our discussion to solutions for the
neutron star spin history with a time-dependent mass inflow rate
satisfying $\gamma (t)= \gamma_0 (t/t_0)^{-2}$. Let
$w \, = \, {1/{\Omega^2 t}}$,
$a=2\beta /I$ and $b=2\gamma_0 t_0^2/I$, then the
differential equation (\ref{eq:rotloss}) becomes
\begin{equation}
w \, + \, t \, {{dw}\over{dt}} \, = \, a \, + \, b w^2
\end{equation}
This can be separated and integrated:
\begin{equation}
\int {{dw}\over{bw^2 \, - \, w \, + \, a}} \, - \, \ln t \, = \,
{\rm constant}
\end{equation}

If $4ab < 1$ then the integration yields
\begin{equation}
{1\over c} \, \ln \left | {{2bw \, - \, 1 \, - \, c} \over
{2bw \, - \, 1 \, + \, c}} \right |
\, - \, \ln t \, = \, {\rm constant}
\end{equation}
where
$c \, = \, \sqrt{1 \, - \, 4ab}\;$.
The constant is fixed by the value of the variable $w$ at some initial
time $t_i$. The solution for $\Omega$ can then be easily derived:
\begin{equation}
\Omega^2 = \frac{2b}{t}\;
\frac{(2bw_i-1+c)-(2bw_i-1-c)(t/t_i)^c}{(1+c)(2bw_i-1+c)-
(1-c)(2bw_i-1-c)(t/t_i)^c}\;.
\label{eq:omega}
\end{equation} 
At later times, $t\gg t_i$, and the solution can be well approximated by
\begin{equation}
\Omega^2\approx \frac{1+c}{2at}\;. 
\label{eq:omega2}
\end{equation}
Evaluating the above equation at the present time $t_0$ yields
\begin{equation}
t_0=\frac{I}{2 \beta \Omega_0^2 \left( 1 + \frac{\gamma_0}{\beta
\Omega_0^4}\right)}.
\label{eq:trueageunsteady}
\end{equation}
Note that the factor $I/(2 \beta \Omega_0^2)$ is the pulsar timing age
(in the case of pure magnetic dipole radiation losses).  Although this
solution corresponds to rather simple evolution laws for $\Omega$ and
$\gamma$, it is useful as an illustration of the effect of allowing
for a time-varying mass inflow rate in the equations (as compared to
the steady case considered in \S2.1). Given that a fallback disk
should have a mass inflow rate decreasing approximately as $t^{-1.2}$
(Cannizzo et al. 1990; Chatterjee et al. 2000; Menou et al. 2001), the
steady and simple time-varying cases considered here may be expected
to bracket the real case ($\gamma \propto t^{-1.2}$), for which a
numerical integration is required in full generality.

\section{Applications}

\subsection{The Crab Pulsar}

For the Crab pulsar (PSR B0531+21), we adopt $I \approx
10^{45}$~g~cm$^{-2}$ and $\beta = 3.85 \times 10^{29}$~erg~s$^{3}$.
These values are derived from measurements of $\Omega$, $\dot \Omega$,
the estimated value of $- \dot E \approx 5 \times
10^{38}$~erg~s$^{-1}$ and the assumption of pure magnetic dipole
radiation losses (see Shapiro \& Teukolsky 1983; Taylor, Manchester \&
Lyne 1993). The propeller losses are comparatively small in the
applications considered below and for the Crab pulsar:
\begin{equation}
\frac{\beta \Omega^4}{\gamma} \approx 25 ~\left(
\frac{\dot M}{10^{16}~{\rm g~s^{-1}}}\right)^{-1}.
\end{equation}
We recall here that, in 1972, the true age of the Crab pulsar was
918~yrs, while its timing age was measured to be 1243~yrs (which is
half the characteristic age as defined by Shapiro \& Teukolsky
1983). The braking index of the Crab pulsar is $n=2.509$ (e.g. Lyne et
al. 1988).

In the case of steady mass inflow in the fallback disk outside the light
cylinder, we find using Eq.~(\ref{eq:nsteady}) that the braking index
is $n =$~2.85, 2.57 and 1.86 for the mass inflow rates $\dot M=
10^{16}$, $3 \times 10^{16}$ and $10^{17}$~g~s$^{-1}$, respectively.
For the same mass inflow rates, Eq.~(\ref{eq:trueagesteady}) predicts a
system age of $t_0 =$~1130, 1101 and 1018~yrs, respectively (in the
limit of very large initial angular speed $\Omega_i$).

In the case of a mass inflow rate decreasing with time as $t^{-1.2}$,
we find using Eq.~(\ref{eq:nunsteady}) that the braking index is $n
=$~2.97, 2.93 and 2.78 for the (current) mass inflow rates $\dot M=
10^{16}$, $3 \times 10^{16}$ and $10^{17}$~g~s$^{-1}$, respectively.
The age cannot be simply computed with the power law index $\alpha = 1.2$.
However, it will be bracketed by the values obtained 
for steady mass inflow above and the $\alpha=2$ case, for which
Eq.~(\ref{eq:trueageunsteady}) holds.  For the same (current)
mass inflow rates as above, Eq.~(\ref{eq:trueageunsteady}) predicts a
system age of $t_0 =$~1099, 1020 and 817~yrs, respectively.

In each of the cases above (steady mass inflow and $\gamma \propto
t^{-2}$), it is possible to find a mass inflow rate for which both the
Crab pulsar true age and braking index are predicted to within
$20\%$ or better. Consequently, it appears that a fallback disk with a current
mass inflow rate in the range $3 \times 10^{16} - 10^{17}$~g~s$^{-1}$
can account for the properties of the Crab pulsar with an accuracy of $20\%$ or so.

The model can be further tested with the third time derivative of the
Crab pulsar frequency (Lyne et al. 1988; Blandford \& Romani 1988). It
is useful to define a second deceleration parameter:
\begin{equation}
p \equiv \frac{ \stackrel{\cdots}{\Omega} \Omega^2}{\dot \Omega^3} = 6 +
\left( 3 - \frac{1}{\frac{1}{4} + \frac{\beta \Omega^4}{4
\gamma}}\right)^2 -\frac{1+\frac{\dot \gamma}{4 \gamma}
\frac{\Omega}{\dot \Omega}}{\frac{1}{4} + \frac{\beta \Omega^4}{4
\gamma}}+ \frac{\frac{\ddot \gamma}{ \gamma} 
 \frac{\Omega^2}{\dot \Omega^2} - \frac{\dot
\gamma}{\gamma} \frac{\Omega}{\dot \Omega}}{1 + \frac{\beta
\Omega^4}{\gamma}},
\end{equation}
which has the correct limit $p=15$ for $\gamma \rightarrow 0$ (pure
magnetic dipole radiation). The expression for steady mass inflow is
found by forcing $\ddot \gamma=\dot \gamma =0$ in the above equation.
Note again that $\alpha(\alpha+1)\gamma/\ddot \gamma $ is the square
of the pulsar true age for a power law evolution of the disk mass
inflow rate ($\gamma(t) \propto t^{-\alpha}$), while $-\Omega / \dot
\Omega$ is twice the pulsar timing age. The measured value of $p$ for
the Crab is $\approx 10.2$ (Lyne et al. 1988; Blandford \& Romani
1988; Melatos 1997). We find, for the case of steady mass inflow, that
the equation above predicts $p=12.6$ for $\dot M= 3 \times
10^{16}$~g~s$^{-1}$ and $p=9.46$ for $\dot M= 10^{17}$~g~s$^{-1}$. For
a time dependent evolution with $\alpha=1.2$, we find $p=13.6$ and
$p=11.96$ for the same mass inflow rates, respectively.  The second
deceleration parameter is therefore correctly predicted by the model
(with an accuracy of $30\%$ or so) for fallback disk mass inflow rates
in the range $3 \times 10^{16} - 10^{17}$~g~s$^{-1}$.

\subsection{Other Young Radio Pulsars}

Braking indices have been measured for four other young radio
pulsars. The measured values are $n=2.2 \pm 0.1$, $1.4 \pm 0.2$, $2.91
\pm 0.05$ and $2.837 \pm 0.001$ for PSR B0540-69, PSR B0833-45 (Vela),
PSR J1119-6127 and PSR B1509-58, respectively (Lyne, Pritchard \&
Graham-Smith 1993; Deeter, Nagase \& Boynton 1999; Lyne et al. 1996;
Kaspi et al. 1994; Camilo et al. 2001).

It is possible to calculate, within the framework of our pulsar
spindown model, the disk mass inflow rates required in each system to
account for the measured breaking indices. In the absence of known
true ages for these four pulsars, we calculate theoretical braking
indices in two limits: (1) the steady mass inflow case
(Eq.~\ref{eq:nsteady}) and (2) the time-dependent case
(Eq.~\ref{eq:nunsteady}) with the additional simplifying assumption
that the pulsar timing age is a correct measure of the pulsar true age
(the validity of this assumption, appropriate for the magnetic dipole
model, was recently challenged by Gaensler \& Frail 2000). The pulsar
properties are taken from Taylor et al. (1993) and Camilo et
al. (2001).

We find that the pulsar spindown model can account for the braking
indices of PSR B0540-69, PSR J1119-6127 and PSR B1509-58 for disk
mass inflow rates $\sim 3 \times 10^{16}$, $\lsim 10^{14}$ and $\lsim
10^{15}$~g~s$^{-1}$, respectively.  In each case, the propeller losses
represent a reasonably small fraction of the total pulsar rotational
losses. The very small value of $n \approx 1.4$ for the Vela pulsar
cannot be accounted for by a model in which propeller losses are
small. We note that the sub-Eddington mass inflow rates required for PSR
B0540-69, PSR J1119-6127 and PSR B1509-58 (as well as the Crab) appear
consistent with the values expected at late stages (before
the fallback disk becomes neutral) in the evolutionary scenario
described by Menou et al. (2001). One expects the magnetospheric
interaction to be largely reduced or even suppressed when the gaseous
disk becomes neutral (independently of the passive-disk outcome
advocated in Menou et al. 2001).

\subsection{Observational Constraints on Disk Sizes}

The presence of disks around pulsars cannot be hidden: even a
relatively compact disk will produce some optical emission, as
radiation at these wavelengths is produced at relatively small
radii.

Optical emission is a common feature of pulsars: since the first
detection of the Crab in 1969, several more pulsars have been observed
in the optical bandpass, and some of them have been seen to pulsate.
In the cases where no pulsations are seen, the optical radiation is
believed to be mainly of thermal origin, while pulsed optical emission
is generally considered to be either a mixture of thermal and
non-thermal (magnetospheric) radiation, or purely
magnetospheric. Pacini (1971) and Pacini \& Salvati (1987) proposed
that the high energy emission comes from relativistic electrons
radiating via synchrotron processes in the outer regions of the
magnetosphere.

The Crab, Vela, and PSR B0540-69 pulsars belong to the above class, all
possessing a pulsating optical counterpart. On the other hand, no
pulsations have been detected for the candidate optical counterpart of
the source PSR B1509-58 proposed\footnote{The
interpretation of this optical counterpart is however a bit
controversial (Mignani et al. 1998; Chakrabarty
\& Kaspi 1998; Gvaramadze 2001).} by Caraveo et al. (1994).  

In the following, we estimate, for each of the sources that we
consider (except for PSR J1119-6127, for which no optical limits are
available so far), the maximum size allowed for a disk to be
compatible with the observed optical limits.  For the three sources
for which the optical emission is pulsed, we consider as an upper
limit the lowest value of the emission (i.e. the minimum of the light
curve), while for PSR B1509-58 we consider the detected value,
assuming that it is indeed associated with the pulsar.

The inner radius of the disk is assumed to coincide with the light
cylinder radius, $R_{lc}$.  The disk model that we adopt is similar to
that described by Perna, Hernquist \& Narayan (2000). We include both
the flux due to viscous dissipation and that due to reradiation. The
irradiating X-ray luminosity is assumed to be isotropic, because, due
to light bending in the vicinity of the star, beaming effects are
smoothed out (Perna \& Hernquist 2000).

Using the X-ray luminosities and periods from observations (see Becker
\& Trumper 1997 for a review), we find (assuming a disk inclination
angle of $60^o$) that, for a mass inflow rate $\sim 10^{17}$ g
s$^{-1}$ a disk of radial extent $\la 6\times 10^9$ cm is compatible
with the optical limits for the Crab pulsar. A large disk ($\ga
10^{13}$ cm) is allowed for both PSR B0540-69 and PSR B1509-58 at the
mass inflow rate of $\sim 3\times 10^{16}$ and $\sim 10^{15}$ g
s$^{-1}$ respectively, while for Vela only a very small disk ($\la
10^{9}$ cm) at an high inclination ($\ga 89^o$) would be compatible
with the observational constraints\footnote{Note however that, as the
inner radii are $ \sim 10^8-10^9$~cm for all the pulsars considered,
we expect most of the intrinsic light from the disk to be produced in
the UV, which is difficult to detect.}.  The allowed size becomes
larger as $\dot{M}$ decreases.  Again, we note that, with the
exception of the Vela pulsar, these disk sizes appear consistent with
the values expected at late stages (before the fallback disk becomes
neutral) in the evolutionary scenario of Menou et al. (2001).

Support for this work was provided by NASA through Chandra Fellowship
grant PF9-10006 awarded by the Smithsonian Astrophysical Observatory
for NASA under contract NAS8-39073.

\end{document}